\documentclass[prd, onecolumn, nofootinbib, floatfix]{revtex4}

\usepackage{amsmath}
\usepackage{graphicx}
\usepackage{color}
\begin{document} 

\title{Prospects for constraining the shape of non-Gaussianity with the scale-dependent bias}

\author{Jorge Nore\~na$^1$, Licia Verde$^{1, 2}$, Gabriela Barenboim$^3$, Cristian Bosch$^3$
\vspace{0.1cm}}

\affiliation{$^1$Instituto de Ciencias del Cosmos, University of Barcelona (ICC-UB/IEEC), Marti i Franques 1, Barcelona 08028, Spain.
\\
$^2$ICREA, Instituci\'o Catalana de Recerca i Estudis Avan\c{c}ats.
\\
$^3$Departament de Fisica Teorica and IFIC, Universitat de Valencia-CSIC, E-46100,
Burjassot, Spain.}

\begin{abstract}

We  consider whether  the  non-Gaussian scale-dependent halo bias  can be used not only to constrain the local form of non-Gaussianity but also to  distinguish among different shapes. In particular, we ask whether it can constrain the behavior of the primordial three-point function in the squeezed limit where one of the momenta is much smaller than the other two. This is potentially interesting since the observation of a three-point function with a squeezed limit that does not go like the local nor equilateral templates would be a signal of non-trivial dynamics during inflation. To this end we use the quasi-single field inflation model of Chen \& Wang \cite{Chen:2009zp,Chen:2009we} as a representative two-parameter model, where one parameter governs the amplitude of non-Gaussianity and the other the shape. We also perform a model-independent analysis by parametrizing the scale-dependent bias as a power-law on large scales, where the power is to be constrained from observations. We find that proposed large-scale structure surveys (with characteristics similar to the dark energy task force stage IV surveys) have the potential to distinguish among the squeezed limit behavior of different bispectrum shapes for a wide range of fiducial model parameters. Thus the halo bias can help discriminate between different models of inflation.

\end{abstract}

\date{\today}

\maketitle 

\section{Introduction\label{sec:intro}}

The study of the deviation from Gaussianity of the initial conditions set by inflation is one of the most active fields of research in cosmology today. This is for a good reason since it is a potential observational handle on the interactions of the inflaton. Indeed, a free field is Gaussian and all its correlation functions are fixed by Wick's theorem in terms of its two-point correlation function 
\begin{equation}
\langle \zeta(\vec{k}_1) \zeta(\vec{k}_2) \rangle = (2\pi)^3 \delta(\vec{k}_1 + \vec{k}_2) P_\zeta (k)\;,
\end{equation}
where $\zeta$ is the comoving curvature perturbation, and $P_\zeta(k) = 2\pi^2\Delta_\zeta^2 (k/k_p)^{(n_s - 1)} (1/k^3)$. The spectral index $n_s$ and amplitude of the scalar power spectrum $\Delta_\zeta^2$ have been measured by the WMAP satellite \cite{Komatsu:2010fb} to be $n_s = 0.968 \pm 0.012$ and $\Delta_\zeta^2 = (2.430 \pm 0.091)\times10^{-9}$ at the $68\%$ confidence level, and $k_p$ is an arbitrary pivot scale there chosen to be $k_p = 0.002\,\mathrm{Mpc}^{-1}$. The interactions of the field induce non-trivial higher-order correlation functions which if observed can serve to discriminate between different models of inflation. In the context of cosmology, Non-Gaussianity refers to the study of the generation of these higher-order correlation functions in different models of the early universe, their modification through the non-linear evolution of the later universe, and how observations may be used to measure them. Of these, the three-point function is expected to be the most accessible to observations; it can be written as
\begin{equation}
\langle \zeta(\vec{k}_1) \zeta(\vec{k}_2) \zeta(\vec{k}_3)\rangle = (2\pi)^3 \delta(\vec{k}_1 + \vec{k}_2 + \vec{k}_3) B_\zeta (k_1, k_2, k_3)\;,
\end{equation}
where $B_\zeta$ is called the bispectrum. Early work focused on the study of the three-point function given by the following phenomenological prescription
\begin{equation}
\zeta = \zeta_g + \frac{3}{5}f^{\rm Loc}_{\rm NL}(\zeta_g^2 - \langle\zeta_g^2\rangle)\;,
\end{equation}
where $\zeta_g$ is a Gaussian random field. This generates a bispectrum with the following shape
\begin{equation}
B_\zeta^{\mathrm{loc}}(k_1, k_2, k_3) = \frac{6}{5}f^{\rm Loc}_{\rm NL}\big[P_\zeta(k_1)P_\zeta(k_2) + \mathrm{cycl.}\big]\;.
\label{eq:local}
\end{equation}
This type of bispectrum is called local since it stems from a local redefinition of the field. Note that it is divergent in the limit in which one of the momenta is much smaller than the other two, often called the squeezed limit; this divergence goes as $\langle\zeta(\vec{k}_1) \zeta(\vec{k}_2) \zeta(\vec{q})\rangle \overset{q\rightarrow 0}{\sim} 1/q^3$.

The three-point function generated by the simplest single-field slow-roll models with a standard kinetic term was first computed by \cite{Maldacena:2002vr,Acquaviva:2002ud}, who found it to be too small to be accessible to observations in the foreseeable future. A ``large'' non-Gaussianity can be generated by introducing new ingredients in the model such as additional light fields (see e.g. \cite{Bartolo:2003jx}), a non-trivial kinetic term (see e.g. \cite{Creminelli:2003iq}), or features in the power spectrum (see e.g. \cite{Flauger:2010ja}). Each of these models generates a characteristic type of non-Gaussianity. Furthermore, within the context of single-field slow-roll inflation, the observation of the three-point function could potentially fix the coefficients of the leading terms in the effective Lagrangian for the perturbations of the inflaton \cite{Cheung:2007st,Senatore:2009gt}.

The best constraints on the primordial three-point function to date come from the WMAP satellite observing the cosmic microwave background radiation (CMB) \cite{Komatsu:2010fb}. A non-trivial three-point function has yet to be observed, but the Planck satellite promises to greatly improve on current bounds \cite{:2006uk}. Another promising probe comes from the observation of large scale structure in the universe (LSS). Competitive constraints on the non-Gaussianity have been found \cite{Slosar:2008hx} using the fact that a non-Gaussianity of the local type, Eq. \eqref{eq:local}, induces a characteristic scale, redshift and mass dependence on the halo bias \cite{Dalal:2007cu,Matarrese:2008nc,Afshordi:2008ru}.

In this paper we wish to study the power of LSS surveys in the near future to constrain the shape of non-Gaussianity through the observation of the halo bias. The halo bias is sensitive to a configuration in which one of the three Fourier modes is much smaller than the other two, the so-called squeezed limit, where the long mode is the scale of observation. The local non-Gaussianity has a divergent squeezed limit and gives a contribution to the large-scale halo bias that goes like $\sim 1/k^2$. In section \ref{sec:model} we summarize the information about inflation contained in the squeezed limit of the bispectrum. It has been shown \cite{Maldacena:2002vr,Creminelli:2004yq,Creminelli:2011rh,Creminelli:2012ed} that the bispectrum of single-field inflationary models is suppressed in the squeezed limit by two powers of the small momentum, thus giving a negligible contribution to the halo bias unless there are features in the power spectrum close to the relevant scales. Models with several light fields may generate a bispectrum which has a squeezed limit that behaves like the local non-Gaussianity thus giving potentially interesting effects to the halo bias in future surveys, as studied for example in \cite{Carbone:2008iz}.
In order for the three-point function to have a non-standard behavior in the squeezed limit, some non-trivial dynamics must be present during inflation. One such model is the ``quasi single-field'' inflation model of \cite{Chen:2009zp,Chen:2009we}, which is the one we use as an example of the power of a future survey to constrain the behavior of the three-point function in the squeezed limit. It consists of a light inflaton coupled to an isocurvaton with a mass which is close to the Hubble scale during inflation. In section \ref{sec:bias} we summarize how the presence of primordial non-Gaussianity affects the halo bias in the peak background split formalism. In section \ref{sec:setup} we describe the survey setup we consider, with the characteristics of the specific Dark Energy Task Force stage IV survey. In particular, we use a survey similar to that used in \cite{Fedeli:2010ud} and we refer the reader to the survey description given on that reference. In section \ref{sec:forecast} we present our main result, namely the forecast of the constraints that such a survey would be able to impose on the parameters of the quasi-single field inflation model, which can also be thought as a parametrization of the squeezed limit of a three-point function. We  also perform the same analysis for a power-law parametrization of the non-Gaussian bias on large scales. Finally, in section \ref{sec:discussion} we draw conclusions and discuss the results.

\section{Model of non-Gaussianity\label{sec:model}}

As we will review in the following section, the non-Gaussian modification of the halo bias is mainly sensitive to the squeezed limit of the bispectrum in which one of the momenta, here denoted by $q$, is much smaller than the other two. It is thus of interest to consider what information can be gained about inflation from constraints on the squeezed limit of the three-point function.

Generic single-field models of inflation must satisfy the following consistency relation
\begin{equation}
\langle \zeta(\vec{k}_1)\zeta(\vec{k}_2)\zeta(\vec{q}) \rangle \overset{q \rightarrow 0}{=} -(2\pi)^3 \delta(\vec{k}_1 + \vec{k}_2 + \vec{q}) P(k_1) P(q) \bigg[-(n_s - 1) + \mathcal{O}\bigg(\frac{q^2}{k^2}\bigg)\bigg]\;.
\label{eq:consistency}
\end{equation}
This relation was first derived in Refs. \cite{Maldacena:2002vr,Creminelli:2004yq}, while Ref. \cite{Creminelli:2011rh} showed that the corrections are suppressed by the square of the small momentum (\emph{i.e.}, the second term in the square brackets). This implies that all single field models of inflation generate a bispectrum which in the squeezed limit goes like $\langle \zeta^3 \rangle \sim 1/q$ (\footnote{In order to prove this, one must assume that the time variation of $\zeta$ outside of the horizon is suppressed at least as $q^2$. This is the case for most models of inflation present in the literature.}). This was understood in reference \cite{Creminelli:2012ed} to be a consequence of the fact that, under the assumption of adiabaticity, correlation functions of $\zeta$ are endowed with a $SO(4,1)$ symmetry which is non-linearly realized by $\zeta$.

The exact momentum dependence of the bispectra as predicted by different models of inflation is too cumbersome to be used in CMB data analysis. What is often done is to look for a bispectrum shape, often called a \emph{template}, which has a high overlap with the exact bispectrum and use it to analyze the data. Two such templates which are often used are the equilateral one \cite{Babich:2004gb} and the orthogonal one \cite{Senatore:2009gt}. They are similar to shapes generated by different operators in the effective Lagrangian for single field inflation \cite{Cheung:2007st,Senatore:2009gt}. However, one must be careful when using these templates with the scale-dependent halo bias to put constraints on the models which inspired them. Indeed, the orthogonal template has a squeezed limit going like $\langle \zeta^3 \rangle \sim 1/q^2$, which is in contradiction with the above-mentioned fact that all single field models of inflation must have a squeezed limit going like $\langle \zeta^3 \rangle \sim 1/q\;$(\footnote{ It is worth noting however that Reference \cite{Senatore:2009gt} gives in their Appendix a template for the orthogonal shape which does have the correct behavior in the squeezed limit.}). One can still view studies using the orthogonal template and the halo bias as examples of the usefulness of this method in constraining intermediate shapes. Nevertheless, at the time of writing, no model in the literature generates a bispectrum which has both a large overlap with and a squeezed limit like that of the orthogonal template. 

For models with several light fields which can contribute to the generation of primordial perturbations none of the above arguments hold, and the bispectrum can generically be large in the squeezed limit. For example, curvaton models (see e.g. \cite{Bartolo:2003jx}) predict a bispectrum which is very close to having a shape like that of the local template, and which goes like $\langle \zeta^3 \rangle \sim 1/q^3$ in the squeezed limit. This is one of the reasons why observing a large non-Gaussianity of the local type is compelling, as it would rule out all single-field models of inflation. However, there is still more information contained in the behavior of the bispectrum in the squeezed limit. Reference \cite{Creminelli:2011rh} argues that in multi-field models, when all the fields are much lighter than the Hubble scale during inflation and are thus scale invariant, the squeezed limit of the bispectrum is proportional to the power spectrum of the long mode, \emph{i.e.}, $\propto 1/q^3$, with corrections which are again quadratic in the small momentum. Thus, in single-field models of inflation and in models with multiple light fields the bispectrum always goes as $1/q^3$ or $1/q$ in the squeezed limit.

If one wishes to obtain a non-standard behavior in the squeezed limit, the alternative is to consider a multi-field model where at least one of the fields has a mass close to $H$, which is high enough to invalidate the arguments of the previous paragraph, but low enough so that it can't be trivially integrated out. An interesting realization of this is the model of Ref. \cite{Chen:2009zp,Chen:2009we}, termed ``quasi-single field inflation''. It consists of two fields, one being a light inflaton and the other an isocurvaton with a mass of order $H$. The inflaton and isocurvaton are coupled through a turning trajectory in field space, which allows for the isocurvaton to have a sizeable contribution to the three-point function of $\zeta$. Reference \cite{Chen:2009zp} computes the bispectrum for such a model, and suggest the following template as a good approximate description
\begin{equation}
B_\zeta(k_1, k_2, k_3) = (2\pi)^4\Delta_\zeta^4 k_p^{2(1 - n_s)}\frac{3^{7/2}}{10 N_\nu(\alpha/27)}\frac{f_{\rm NL}}{(k_1 k_2 k_3)^{3/2} (k_1 + k_2 + k_3)^{3/2}} N_\nu \bigg(\frac{\alpha k_1 k_2 k_3}{(k_1 + k_2 + k_3)^3}\bigg)\;,
\label{eq:template}
\end{equation}
where $N_\nu$ is the Neumann function, $\alpha$ is fixed to be $8$, and $\nu \equiv \sqrt{9/4 - m^2/H^2}$, with $m$ being the mass of the heavy isocurvaton. The factors in front fix the normalization of the template to be equal to the local template at the equilateral point $k_1 = k_2 = k_3$. The amplitude $f_{\rm NL}$ is related to $\nu$, the third derivative of the isocurvaton potential, and the angular velocity in the turning trajectory $\dot\theta_0$ as $f_{\rm NL} = \alpha(\nu)P_\zeta^{-1/2}(\dot\theta_0/H)^3(-V'''/H)$, where $\alpha(\nu)$ is the function plotted in figure 8 of Ref. \cite{Chen:2009zp}. %In the squeezed limit the $f_{\rm NL}$ normalization corresponds to roughly 1/6 of the local $f^{\rm Loc}_{\rm NL}$.

In the squeezed limit, the template of Eq. \eqref{eq:template} goes like $\langle \zeta^3 \rangle \sim 1/q^{3/2 + \nu}$; thus different values of $\nu$ give different behaviors in the squeezed limit. For $\nu = 1.5$, the template behaves like the local one in the squeezed limit; for $\nu = 0.5$ it behaves like the orthogonal \emph{template} in the squeezed limit;  other values of $\nu$ give intermediate behaviors. One therefore expects future surveys to constrain the value of $\nu$ using the halo bias.

The halo bispectrum is actually not sensitive to the exact squeezed limit in which one of the Fourier modes is infinitely much smaller than the other two. This means that even if all single field models must satisfy the consistency relation of equation \eqref{eq:consistency} in the exact squeezed limit, there might be single field models with features in the bispectrum close to the actual scales relevant for observations such that they can still produce a large effect in the halo bias. Some work in this direction has been done in Ref. \cite{Huang:2012mr}. Another possibility is that the bispectrum breaks scale invariance, again in this case the model must satisfy the consistency relation of equation \eqref{eq:consistency} but can still have an effect on the halo bias, see Ref. \cite{Sefusatti:2009xu}. Refs. \cite{Ganc:2012ae,Agullo:2012cs} very recently argued that a modified vacuum state can have a feature near to the squeezed limit that induces a scale dependence of the halo bias which is stronger than that generated by the local template.

In this paper we wish to forecast the constraints that a future survey with the characteristics of the Dark Energy Task Force stage IV \cite{Albrecht:2006um} can put simultaneously on $\nu$ and $f_{\rm NL}$ as defined in equation \eqref{eq:template}. In the context of the quasi-single field model, a constraint on $\nu$ translates into a constraint on the mass of the heavy isocurvaton. More generally, one can see this as an estimate of the power of future surveys to constrain the behavior of the squeezed limit of the three-point function. In section \ref{sec:bias} we also introduce a parametrization of the non-gaussian modification to the halo bias on large scales in order to perform this analysis in a model independent way. All of this is potentially interesting since if a bispectrum with a squeezed limit that does not go like $1/q$ or $1/q^3$ is observed, it would signal the presence of heavy fields or other non-trivial dynamics during inflation.

\begin{figure}[h]
\begin{center}
\includegraphics[scale=1.0]{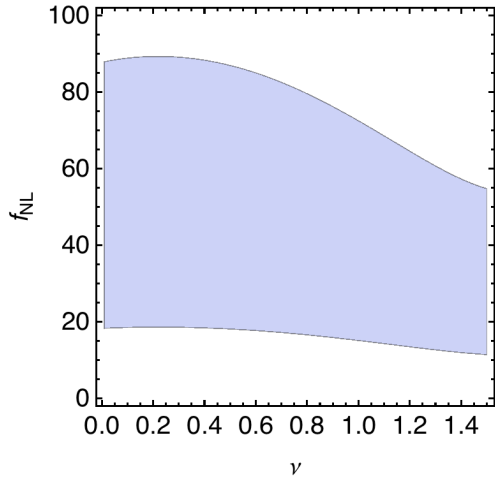} 
\end{center}
\caption{Estimated WMAP 7 constraints on $f_{\rm NL}$ at $1$--$\sigma$ coming from the overlap between the quasi-single field inflation template of Eq. \eqref{eq:template} and the local template of Eq. \eqref{eq:local} as defined in Eq. \eqref{eq:cosine}. }
\label{fig:cmb}
\end{figure}

Let us now estimate the values of $f_{\rm NL}$ that are allowed by current observations. Though no dedicated data analysis has been performed for the template of Eq. \eqref{eq:template}, it has a sizable overlap with the local template. To be more quantitative, define the ``cosine'' between two bispectra as
\begin{equation}
\cos(B_1, B_2) \equiv \frac{B_1\cdot B_2}{\sqrt{B_1\cdot B_1 B_2 \cdot B_2}} \qquad \mathrm{where} \quad B_1 \cdot B_2 \equiv \sum_{k_1,k_2,k_3} B_1(k_1, k_2, k_3) B_2(k_1, k_2, k_3)\;.
\label{eq:cosine}
\end{equation}
One can then use current constraints on the local, equilateral and orthogonal templates \cite{Komatsu:2010fb} to estimate the values of $f_{\rm NL}$ compatible with current data analyses:
\begin{equation}
f_{\rm NL} \approx f_{\rm NL}^{\rm Loc} \cos(B_\zeta, B_{loc}) + \dots\;,
\end{equation}
where the dots indicate other templates which have been analyzed; we don't use them here since the tightest constraints come from those on the local template. Indeed, the cosine between the template under study, Eq. \eqref{eq:template}, and the local template, Eq. \eqref{eq:local}, is always $\gtrsim 0.6$. We plot the estimated $1$--$\sigma$ constraints in figure \ref{fig:cmb}. Note that this is only a rough estimate, since the true constraints coming from the CMB data would require a dedicated data analysis, which is beyond the scope of this article. One can also improve on this estimate by using a spherical cosine rather than the flat one defined in Eq. \eqref{eq:cosine}; we expect the results to be of the same order of magnitude as those plotted in figure \ref{fig:cmb}.

\section{Modeling of the non-Gaussian halo bias\label{sec:bias}}

In this section we wish to summarize the computation of the modification of the halo power spectrum in the presence of non-Gaussian initial conditions. Let us begin by introducing some notation.

The comoving gauge curvature perturbation $\zeta$, in terms of which the initial conditions set by inflation are often expressed, is related to the linear dark matter density contrast $\delta$ at redshift $z$ through the transfer function $T$ and the linear growth factor $D$ normalized to one at redshift zero
\begin{equation}
\delta(k) = k^2 T(k) D(z) \zeta(k)\;,
\label{eq:transfer}
\end{equation}
where we have written this equation in Fourier space. These factors contain the evolution of the gravitational potentials once the perturbation enters the horizon\footnote{Note that a different definition of the transfer function is often used, such that it is normalized to one on large scales. We do not normalize to one, but rather use equation \eqref{eq:transfer} as the definition, which is the one that appears directly in the output file of CAMB at redshift zero (though one must be careful with the factors of $h$).}. The transfer function as written here can be readily computed using publicly available numerical codes like CAMB \cite{Lewis:1999bs}.

In the high-peak formalism, one considers a halo to form when a region of space has an average density contrast larger than some threshold $\delta_c$. Let us define this average as
\begin{equation}
\delta_R = \int \frac{\mathrm{d}^3 k}{(2\pi)^3} W(kR) \delta(k)\,,
\label{eq:filtered}
\end{equation}
where $W$ is a window function that averages out modes with a wavelength larger than the region under consideration. We will take this filter to be a top-hat in real space, which in Fourier space takes the form
\begin{equation}
W(kR) = \frac{3}{(kR)^3}(\sin kR - kR \cos kR)\,.
\end{equation}
The critical density contrast threshold $\delta_c$ is often estimated by computing the time at which a perfectly spherical halo completely collapses and extrapolating the linear density to that point. In an Einstein-de Sitter space one obtains $\delta_c = 1.686$, while in $\Lambda$CDM this estimate receives a small correction and is slightly redshift dependent. The scale $R$ over which one averages is related to the mass of the halo formed by $M = 4\pi/3 \Omega_m \rho_c R^3$, where $\rho_c$ is the critical density of the universe $\rho_c = 3H_0^2/8\pi G$. Equations \eqref{eq:transfer} and \eqref{eq:filtered} suggest the following definition
\begin{equation}
\mathcal{M}_M(k) \equiv k^2 T(k) W(kR)\;,
\end{equation}
where $R$ is expressed in terms of $M$ as above. Thus for example the variance of the filtered density contrast can be computed as
\begin{equation}
\sigma_M^2 = \int \frac{\mathrm{d}^3 k}{(2\pi)^3} \mathcal{M}_M^2(k) P_\zeta(k)\;.
\end{equation}

In the presence of non-Gaussian initial conditions, the (Eulerian) halo power spectrum $P(k)$ is modified
\begin{equation}
P(k) = (b_E^{(g)} + \Delta b )^2 P_\delta(k)\;,
\label{eq:Pk}
\end{equation}
where $g$ denotes the Gaussian case. It was realized in Refs. \cite{Dalal:2007cu,Matarrese:2008nc,Slosar:2008hx,Afshordi:2008ru} that this correction to the power spectrum can have a characteristic scale dependence that makes it relatively easy to observe. One can estimate this correction with an arbitrary bispectrum shape using the peak background split formalism as done originally in Ref. \cite{Matarrese:2008nc}
\begin{equation}
\Delta b(k, M) = \frac{1}{\mathcal{M}_M(k)}\Bigg(\frac{(b_E^{(g)} - 1)\delta_c}{D(z)} \mathcal{F}(k, M) + \frac{\mathrm{d}\mathcal{F}(k,M)}{\mathrm{d}\ln \sigma_M}\Bigg)\;,
\label{eq:deltab}
\end{equation}
where the second term in the parenthesis was computed in Refs. \cite{Desjacques:2011jb,Desjacques:2011mq}, and is due mainly to the modulation of the variance with mass due to the non-Gaussianity. The ``form factor'' $\mathcal{F}$ for an arbitrary bispectrum is defined as
\begin{equation}
\mathcal{F}(k,M) = \frac{1}{8 \pi^2 \sigma_M^2 P_\zeta(k)}\int\mathrm{d}k_1\,k_1^2 \mathcal{M}_M(k_1) \int_{-1}^1\mathrm{d}\mu\,\mathcal{M}_M\big(\sqrt{k^2+k_1^2+2 k_1 k \mu}\big) B_\zeta\big(k,k_1,\sqrt{k^2+k_1^2+2 k_1 k \mu}\big)\;.
\label{eq:F}
\end{equation}
In reference \cite{Desjacques:2011jb} equations \eqref{eq:deltab} and \eqref{eq:F} were found to agree well with simulations \cite{Wagner:2010me} including a local, orthogonal and equilateral templates for the bispectrum of the initial conditions. This agreement has been seen to be good up to a scale of $k \sim 0.1\;h\,\mathrm{Mpc}^{-1}$ \cite{Wagner:2011wx}. 

In the integral of equation \eqref{eq:F}, the scale $k$ corresponds to the scale of observation which for the survey we will study varies between $k \sim 0.003\;h\, \mathrm{Mpc}^{-1}$ and\footnote{In previous analyses that did not consider the correction to the halo bias given by the second term in the right hand side of Eq. \eqref{eq:F}, the maximum $k$ considered was $0.03 h/\mathrm{Mpc}$. This more conservative cutoff ensured that the adopted bias description was correct (on these scales the correction is negligible) but reduced the total signal-to-noise.} $k \sim 0.1\;h\, \mathrm{Mpc}^{-1}$. On the other hand, for a halo mass of order $M = 10^{12}\;M_\odot h^{-1}$, the filter $W(k_1 R)$ suppresses the integral at scales $k_1 \gtrsim 3\;h\, \mathrm{Mpc}^{-1}$, while the transfer function and the factor of $k_1^2$ in the integral suppress the integral at scales $k_1 \lesssim 0.1\;h\, \mathrm{Mpc}^{-1}$. The measure of the integral will thus peak at $k_1 \sim \mathcal{O}(1)\,\mathrm{Mpc}^{-1}$. Therefore the integral will receive a sizable contribution only from the squeezed configuration of the momenta appearing as the arguments of the bispectrum in which $k$ is at least an order of magnitude smaller than $k_1$. We will use equations \eqref{eq:deltab} and \eqref{eq:F} to compute the non-Gaussian modification to the halo bias induced by a template of the form given by equation \eqref{eq:template}.

At very large scales the non-Gaussian modification to the halo bias probes very squeezed configurations. In such a regime $\Delta b(k, M)$ often behaves as a power law in $k$ as can be seen from equations \eqref{eq:deltab} and \eqref{eq:F}. Indeed one expects that for a bispectrum going as $\langle \zeta^3 \rangle \overset{q\rightarrow 0}{\sim} 1/q^a$, the bias behaves as $\Delta b(k, M) \sim 1/k^{a-1}$. In order to perform a model-independent (though perhaps rough) analysis of the power of future surveys to constrain the behavior of the halo bias at large scales, we use the following parametrization (as suggested by \cite{Agullo:2012cs})
\begin{equation}
\Delta b(k, M) = f^{p}_{\rm NL} \frac{A(M)}{k^\beta}\;,
\label{eq:parametrization}
\end{equation}
 where $A(M)$ is a mass-dependent normalization defined such that the bias is the same as that in the presence of a local non-Gaussianity at a scale of $k = 0.03\,h\,\mathrm{Mpc}^{-1}$, and $\beta$ is a parameter to be constrained. Note that $f_{\rm NL}$ of equation \eqref{eq:template} and $f_{\rm NL}^p$ are normalized differently, so that in order to compare the results obtained when using the template of Eq. \eqref{eq:template} and those obtained using the parametrization \eqref{eq:parametrization} one should multiply $f_{\rm NL}^p$ roughly by  a factor of $8$.

\section{Setup\label{sec:setup}}

We wish to estimate the capability of a future survey satisfying the requirements of the Dark Energy Taskforce stage IV \cite{Albrecht:2006um}, to constrain the shape of the bispectrum. For this we assume a survey similar to the one studied in reference \cite{Fedeli:2010ud}. We present the characteristics we assume for such a survey in table \ref{tab:survey}. 
\begin{table}[h]
\begin{tabular}{cc}
\hline
Sky coverage & $2 \times 10^4$ square degrees \\
Minimum redshift & $0.5$ \\
Maximum redshift & $2.1$\\
Typical galaxy halo mass & $10^{12}\;M_\odot h^{-1}$\\
\hline
\end{tabular}
\caption{\label{tab:survey}Description of the Dark Energy Task Force stage IV survey used.}
\end{table}
We also use the results of Ref. \cite{Fedeli:2010ud} for the average number of galaxies available to the survey, the average mass of the galaxies observed, and the Gaussian bias; let us summarize them here. 

For the Gaussian bias we use the results reported in figure 4 of Ref. \cite{Fedeli:2010ud}, who in turn use the results of Ref. \cite{Orsi:2009mj}. There the bias is estimated using the halo model, a semi-analytic model for galaxy formation, and assuming a spectroscopic selection based on $H\alpha$ emission with a threshold given for a survey similar to Euclid. 

The modification of the halo power spectrum due to non-Gaussianity will depend on the halo mass. Here we take the typical masses of the haloes of the galaxies observed to be $10^{12}\;M_\odot h^{-1}$ as in \cite{Fedeli:2010ud}, see their figure 3. We compute the non-Gaussian modification to the halo bias evaluated at this fixed mass. We don't expect the errors induced by fixing the mass in equation \eqref{eq:F} to be of qualitative importance for our results.

The number of galaxies at a given redshift can be computed from the halo mass function $n(z,M)$ (the number of haloes in a differential mass interval at a given redshift), and the first moment of the halo occupation distribution $\langle N_g | M\rangle$ (very roughly the probability that there be $N_g$ galaxies  in a halo of mass $M$)
\begin{equation}
n_g(z) = \int_{M_g}^{\infty} n(M, z) \langle N_g | M \rangle \, \mathrm{d}M\;,
\label{eq:ng}
\end{equation}
where $M_g$ is the minimum halo mass for a galaxy observed by the survey for which we again follow Ref. \cite{Fedeli:2010ud} and take $M_g = 10^{11}\; M_\odot h^{-1}$, see their section 4.4. Since the average number of galaxies is important only in computing the shot noise, which we have verified has a small effect, we neglect the non-Gaussian correction to the mass function and use the Sheth and Tormen mass function \cite{Sheth:2001dp}. As for the first moment of the halo occupation distribution, we follow Ref. \cite{Fedeli:2010ud} and use the following expression \cite{Cooray:2002dia,Cooray:2003dv}
\begin{equation}
\langle N_g | M \rangle = N_{g,0} \bigg(\frac{M}{M_0}\bigg)^\theta\,,
\end{equation}
and the parameters $\theta$, $M_0$ and $N_{g,0}$ depend on the type of galaxy considered. We assume here typical galaxies to be blue, for which $N_{g,0} = 0.7$ and $\theta = 0$ if $M \leq M_0$ and $\theta = 0.8$ otherwise and take $M_0 = 4\times 10^{12}\; M_\odot h^{-1}$ as in section 4.1 of Ref. \cite{Fedeli:2010ud}. Let us stress that the average number of galaxies affects only the shot noise which we verify to be small and our results will not be very sensitive to the assumptions made in computing it.

\section{Forecast\label{sec:forecast}}

We work under the approximation that the Likelihood for the (Eulerian) halo power spectrum is a Gaussian centered around a ``fiducial'' model for which the values of the parameters of interest $f_{\rm NL}$ and $\nu$ are fixed at $\bar{f}_{\rm NL}$ and $\bar{\nu}$
\begin{equation}
\ln \mathcal{L} = -\frac{1}{2} \frac{(\Delta P)^2}{\sigma_P^2}\;,
\end{equation}
where $\Delta P \equiv P - P|_{\bar{f}_{\rm NL},\,\bar{\nu}}$ is the deviation from the fiducial model. The standard procedure is to assume the behavior of the halo power spectrum to be nearly linear on the parameters of interest (here called generically $\theta_i$), so that the Likelihood function is a Gaussian also in terms of those parameters. One can then estimate the variance and covariance of a future survey by computing the Fisher information matrix on the parameters
\begin{equation}
F_{ij} = \frac{\partial^2|\ln\mathcal{L}|}{\partial \theta_i \partial\theta_j}\;.
\end{equation}
This is a good approximation when one is interested in a region in parameter space around the fiducial model which is small with respect to the typical variation of the power spectrum with the parameters, \emph{i.e.}, when the resulting estimated uncertainties are small. However, in the model we are studying the variation of the power spectrum with respect to $\nu$ is highly non-linear and this approximation need not hold. We will instead compute the $\Delta \chi^2$ of each point in parameter space
\begin{equation}
\Delta \chi^2 = \sum_{z,k} \bigg(\frac{\Delta P}{\sigma_P}\bigg)^2 = \sum_{z,k} \bigg(\frac{P}{\sigma_P}\bigg)^2 \frac{\Delta P^2}{P^2}\;,
\label{eq:deltaChiSq}
\end{equation}
where we sum over the different redshift shells and Fourier modes available to the survey. This quantity is in general costly to compute when dealing with a high-dimensional parameter space. However, we will be interested only in varying $f_{\rm NL}$ and $\nu$ since our main focus is the measurement of the non-Gaussianity and its behavior in the squeezed limit. This implies that we keep other cosmological parameters fixed at the WMAP 7 best fit values \cite{Komatsu:2010fb}, and the actual uncertainties on $f_{\rm NL}$ and $\nu$ are expected to be somewhat larger than our estimates. Reference \cite{Carbone:2010sb} estimated this increase in the errors to be mild in the case of a local non-Gaussianity (around $10\% \sim 30\%$ depending on the survey if the number of relativistic neutrino species is kept fixed). This can be understood as being due to the fact that no other cosmological parameter induces a scale dependence and a redshift dependence on the halo bias as the one induced by a local non-Gaussianity. The same argument holds for the range of $\nu$ we will study, and we thus expect the increase in the estimated uncertainty on $f_{\rm NL}$ and $\nu$ due to the uncertainty on other cosmological parameters to be small.

For an infinitesimal shell in Fourier space with a width $\Delta k$ and effective volume $V_{eff}$ the relative error can be estimated by counting the number of Fourier modes
\begin{equation}
\bigg(\frac{\sigma_P}{P}\bigg)^2 = \frac{2}{4\pi k^2 \Delta k V_{eff}/(2\pi)^3}\;.
\end{equation}
The effective volume is the volume corrected by taking into account the shot noise
\begin{equation}
V_{eff} = V(z)\bigg(1 - \frac{1}{n_g(z) P(k)}\bigg)^2\;,
\end{equation}
where $n_g$ is the average number of galaxies, $V$ is the volume of each redshift shell, and we've explicitly stated the dependence on redshift and Fourier mode. Equation \eqref{eq:deltaChiSq} can now be written explicitly
\begin{equation}
\Delta \chi^2 = \sum_{i} \frac{V(z_i)}{(2\pi)^2} \int_{k_{min}}^{k_{max}}\mathrm{d}k\, k^2 \bigg(1 - \frac{1}{n_g(z_i) P(k)}\bigg)^2 \bigg(\frac{\Delta P (k, z_i)}{P(k, z_i)}\bigg)^2\;,
\end{equation}
where the sum is over redshift shells, we take $k_{max} \simeq 0.1\,h / \mathrm{Mpc}$ since it is the scale up to which we trust the calculation of the non-Gaussian modification to the bias \cite{Wagner:2011wx}, and the minimum $k$ is computed as $k_{min} = 2\pi / V(z_i)^{1/3}$. We divide the redshift range in shells with a width $\Delta z = 0.1$. The halo power spectrum and average number of galaxies are given respectively in equations \eqref{eq:Pk} and \eqref{eq:ng}. The deviation of the halo power spectrum from the fiducial model was computed by taking the difference between the halo power spectra of the given model and the fiducial one computed using equation \eqref{eq:deltab} to model the effect of non-Gaussianity.

\begin{figure}[ht!]
\begin{center}
\begin{minipage}{.4\textwidth}
\includegraphics[scale=1.0]{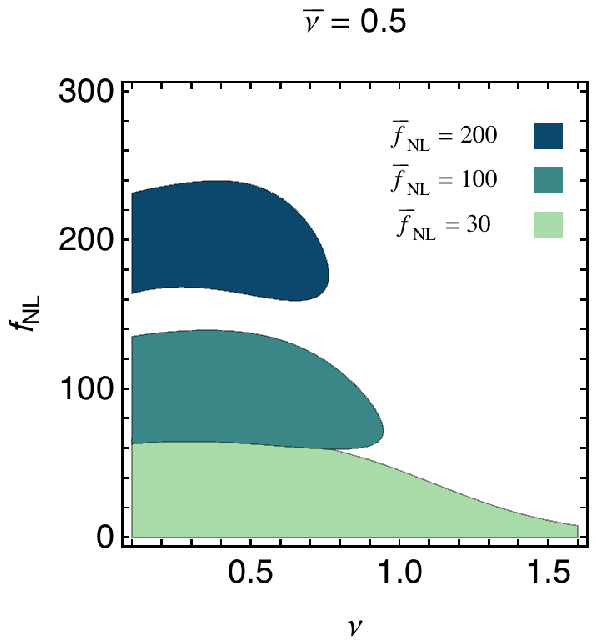} 
\end{minipage}
\hspace{.1\textwidth}
\begin{minipage}{.4\textwidth}
\includegraphics[scale=1.0]{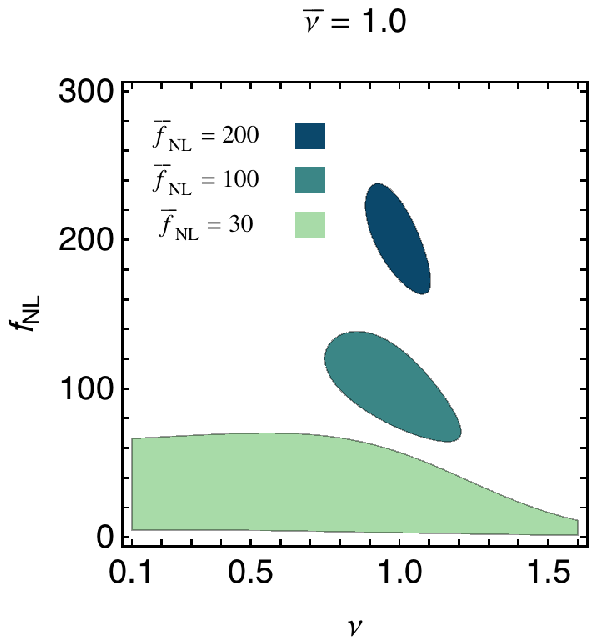} 
\end{minipage}
\vspace{1em}
\begin{minipage}{.4\textwidth}
\includegraphics[scale=1.0]{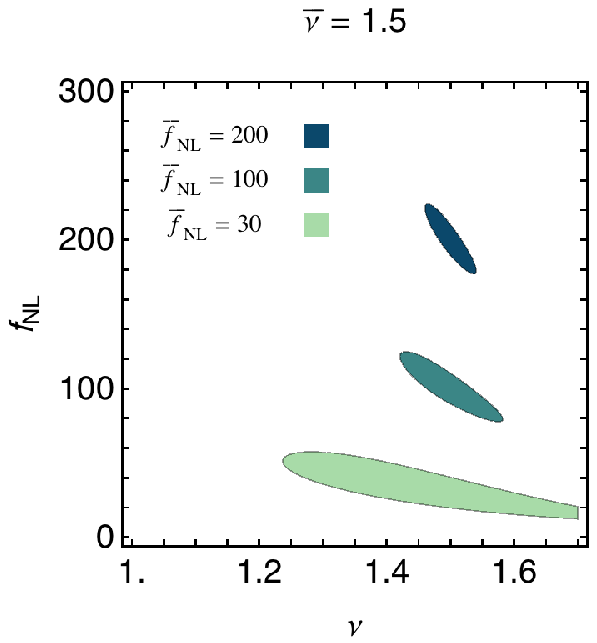} 
\end{minipage}
\end{center}
\caption{Regions in the $f_{\rm NL}$, $\nu$ parameter space defined in Eq. \eqref{eq:template} satisfying $\Delta \chi ^2 \leq 2.3$, corresponding to the $68.3\%$ confidence level. In the squeezed limit this model behaves as $\langle \zeta^3\rangle_{q\rightarrow 0} = 1/q^{\nu + 3/2}$. We show such regions for several fiducial models, showing that the uncertainties decrease as the fiducial value of $f_{\rm NL}$ becomes larger and the fiducial value of $\nu$ approaches $1.5$, which corresponds to a bispectrum shape that behaves like the local template in the squeezed limit.}
\label{fig:chi2}
\end{figure}

In figure \ref{fig:chi2} we show the $\Delta\chi^2 \leq 2.3$ regions corresponding to $68.3\%$ \emph{joint} confidence levels for different values of the fiducial parameters $\bar{\nu}$ and $\bar{f}_{\rm NL}$. For $\bar\nu = 0.5$, which would correspond to a squeezed limit going like $\langle \zeta^3\rangle \overset{q\rightarrow 0}{\sim} 1/q^2$, the constraints on $\nu$ will be broad, and even the detection of non-Gaussianity would be challenging, requiring very large values of $f_{\rm NL}$ in mild tension with current CMB constraints. As $\bar\nu$ increases, the effect on the bias becomes stronger and the survey becomes more sensitive to both $f_{\rm NL}$ and $\nu$. For $\bar\nu = 1.5$, corresponding to a shape that behaves like the local template in the squeezed limit, the survey becomes sensitive to $f_{\rm NL} = \mathcal{O}(10)$, and is able to put constrains on $\bar\nu$. Recall that we plot the 68\% 2-parameter joint confidence regions, a similar confidence interval for a 1-parameter only model --i.e., if $\nu$ were to be kept fixed-- would correspond to $\Delta\chi^2 \leq 1$; thus to compare ours with other analyses where $\nu$ is kept fixed, one can ``slice'' the contours at the preferred $\nu$ value, but should interpret the resulting error as similar to a 90\% confidence region. Note that for large values of $\bar \nu$ and $\bar f_{\rm NL}$ the regions resemble ellipses, signaling the fact that the errors are small relative to the variation of $\nu$ and the Fisher approximation holds. For smaller values of $\bar\nu$ and $\bar{f}_{\rm NL}$ this is not the case and the confidence regions adopt more complex shapes.

\begin{figure}[ht!]
\begin{center}
\begin{minipage}{.4\textwidth}
\includegraphics[scale=1.0]{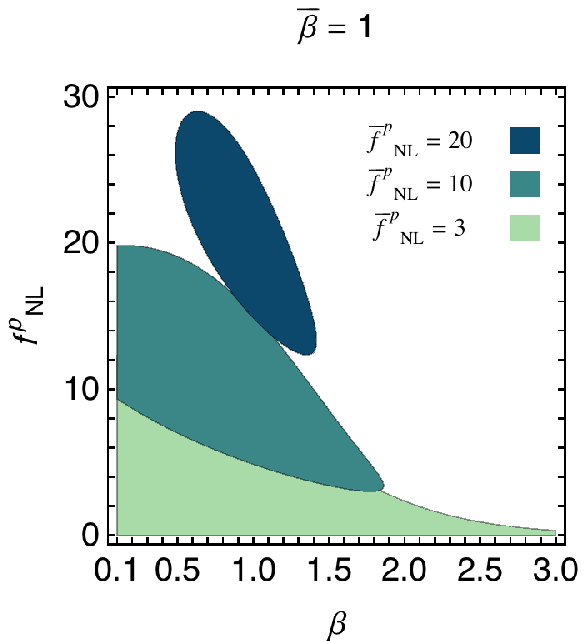} 
\end{minipage}
\hspace{.1\textwidth}
\begin{minipage}{.4\textwidth}
\includegraphics[scale=1.0]{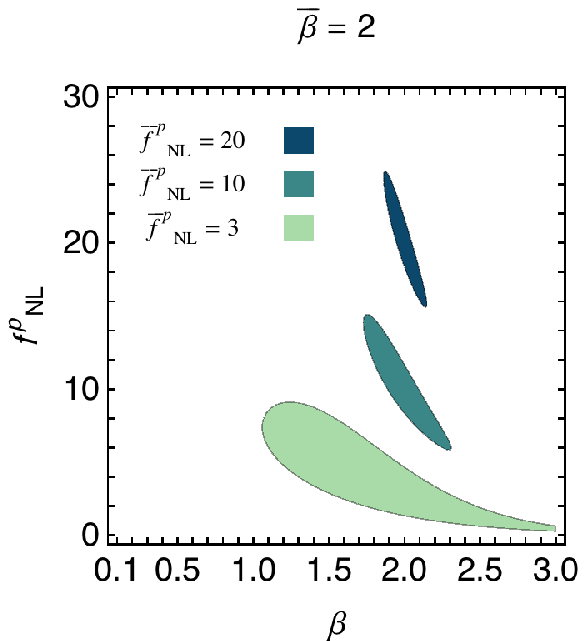} 
\end{minipage}
\vspace{1em}
\begin{minipage}{.4\textwidth}
\includegraphics[scale=1.0]{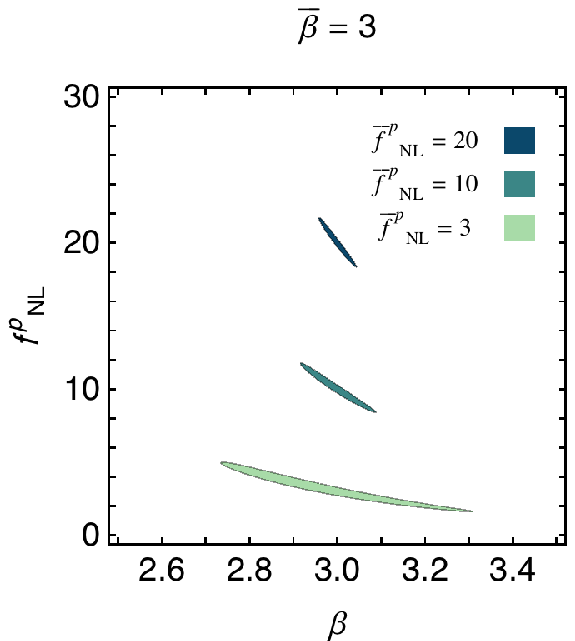} 
\end{minipage}
\end{center}
\caption{Forecasts for the parametrization $\Delta b = f_{\rm NL}^p A / k^\beta$. Green regions correspond to points in the $f_{\rm NL}^p, \beta$ parameter space satisfying $\Delta \chi ^2 \leq 2.3$, corresponding to the $68.3\%$ confidence level. Note that $f_{\rm NL}^p$ is defined such that the non-Gaussian modification to the halo bias, eq. \eqref{eq:deltab}, is the same as in the presence of a local non-Gaussianity at a scale $k = 0.03\,h\,\mathrm{Mpc}^{-1}$, thus $f_{\rm NL}(\nu=1.5)\sim 8f_{\rm NL}^p(\beta=2)$.}
\label{fig:parametrization}
\end{figure}

In figure \ref{fig:parametrization}, we present the same analysis performed for the model-independent parametrization of equation \eqref{eq:parametrization}. Here we restrict ourselves only to large scales where this parametrization can be relevant, and thus integrate up to $k_{max} = 0.03\,h\,\mathrm{Mpc}^{-1}$. The amplitude $f_{\rm NL}$ of the template of Eq. \eqref{eq:template} and $f_{\rm NL}^p$ defined in Eq. \eqref{eq:parametrization} are normalized differently, and one should multiply $f_{\rm NL}^p$ by a factor $\sim 8$ in order to compare the results. Thus, the values of $f_{\rm NL}^p$ in figure \ref{fig:parametrization} are taken to be small for ease of comparison with figure \ref{fig:chi2}. We present the results for $\beta = 1, 2, \mathrm{and}\,\, 3$ corresponding respectively to the behavior of the orthogonal \emph{template}, the local template, and the models studied in Refs. \cite{Ganc:2012ae,Agullo:2012cs}. Again the confidence regions tend to ellipses for larger values of the amplitude $f_{\rm NL}^p$ and power $\beta$, and the results are consistent with those obtained for the quasi-single field inflation model in figure \ref{fig:chi2}.

\section{Summary and discussion\label{sec:discussion}}

We studied the possibility that a future galaxy survey with the characteristics of the Dark Energy Task Force stage IV might be able to constrain the shape of the primordial non-Gaussianity through observations of the halo bias. A non-vanishing primordial three-point function induces a modification of the halo bias that is sensitive mainly to the squeezed limit in which one of the three momenta is smaller than the other two. Single field models produce a bispectrum with a squeezed limit going like $\langle \zeta^3\rangle \sim 1/q$, where $q$ is the small momentum, while models of inflation with multiple light fields can produce a large non-Gaussianity of the local type, \emph{i.e.}, going like $\langle \zeta^3\rangle \sim 1/q^3$ in the squeezed limit. Thus, the observation of a squeezed limit with a behavior which is different to these two possibilities might signal the presence of some non-trivial dynamics during inflation. Such can happen for example in the ``quasi-single field'' inflation model, which consists of a light inflaton and a massive isocurvaton coupled through a turning trajectory. It can produce a shape going as
\begin{equation}
\langle \zeta(\vec{k}_1)\zeta(\vec{k}_2)\zeta(\vec{q})\rangle \overset{q\rightarrow 0}{\sim} 1/q^{3/2 + \nu}\,,
\end{equation}
where $\nu \equiv \sqrt{9/4 - m^2/H^2}$. 

We estimate the ability of a survey such as that described in section \ref{sec:setup} to constrain the squeezed limit of the primordial three-point function. For this purpose we  perform forcasts on the uncertainties on a measurement of the amplitude and shape of the non-Gaussianity generated by the quasi-single field inflation model (see figure \ref{fig:chi2}) and of a phenomenological model-independent parametrization of the non-Gaussian halo bias at large scales $\Delta b = f_{\rm NL}^p A / k^\beta$ (see figure \ref{fig:parametrization}).

Since the variation of the halo power spectrum with the power $\nu$ or $\beta$ is large for small values of the amplitue $f_{\rm NL}$ or $f_{\rm NL}^p$, the Likelihood function in terms of these parameters is not expected to be a Gaussian. We cannot then use the Fisher information matrix approach, and instead compute $\Delta \chi^2$ at each point in parameter space. We report our findings in figures \ref{fig:chi2} and \ref{fig:parametrization}. We find that the forecasted uncertainties depend quite strongly on the fiducial model. If the fiducial model is taken to be similar to a local non-Gaussianity ($\bar \nu = 1.5$ or $\bar\beta = 2$), the survey will be able to put tight constraints on the parameters of the model, even for $\bar{f}_{\rm NL} = \mathcal{O}(10)$. As $\bar \nu$ or $\bar\beta$ decreases the forecasted uncertainties become larger, requiring larger values of $\bar{f}_{\rm NL}$ or $\bar{f}^p_{\rm NL}$ for non-Gaussianity to be observable. This is to be expected since a milder divergence in the squeezed limit translates into a milder scale dependence in the non-Gaussian modification to the halo bias.

Though the most stringent limits on the primordial non-Gaussianity to date come from CMB observations, the CMB has a limited capacity in constraining the shape of the three-point function. Indeed, it is not sensitive to the physically interesting squeezed limit, and would not distinguish between two shapes which have a large cosine but with different squeezed limits. In this sense halo bias observations are complementary to CMB observations and might open a new window to the physics of inflation. It would still be of interest however to find the constraints on the amplitude of the quasi-single field inflation template coming from CMB data.

Another potentially interesting probe of the squeezed limit is the possibility of observing the $\mu$ distortion of the CMB spectrum recently proposed in Ref. \cite{Pajer:2012vz}. A primordial three-point function induces a correlation between the $\mu$ distortion and the temperature which is sensitive to the very squeezed limit. It would be interesting to study the power of such a probe to constrain $f_{\rm NL}$ and $\nu$.

Finally, let us comment on ways in which one can increase the signal to noise ratio in our analysis. If the analysis is repeated for a photometric survey like LSST, which can observe galaxies with fainter magnitudes and thus higher redshift, one expects the gain in volume to reduce the errors by a factor $\sim 2$. Refs. \cite{Seljak:2008xr,Slosar:2008ta,Hamaus:2011dq} argue that with the use of several differently biased tracers and an optimal weighting it is possible to reduce the errors on a local $f_{\rm NL}$ by at least an order of magnitude. We expect a similar reduction to take place in our case, greatly improving the sensitivity of the survey. Another way to reduce the uncertainties is that proposed by Ref. \cite{Cunha:2010zz} who argue in favor of cross-correlating pixels in order to extract more information from the survey; they find that for a local non-Gaussianity the reduction of the uncertainties is again of an order of magnitude. If the uncertainty on $f_{\rm NL}$ can be brought to be of $\mathcal{O}(1)$, general relativistic effects on the halo bias become relevant and might then be observable (e.g., \cite{VM09} and refs therein).

While this paper was being completed, we became aware of a similar analysis being carried out by another group, see Ref. \cite{sefusatti}. We have carried out a rough comparison of our results and find them to be consistent. We are grateful to the authors for coordinating with us the publication on the ArXiv website.

\acknowledgments
We acknowledge stimulating discussions with X. Chen and E. Sefusatti.
J.N. is supported by ERC grant FP7-IDEAS-Phys.LSS 240117. The Work of L.V. is supported by grant FP7-IDEAS-Phys.LSS 240117 and MICINN grant AYA2011-29678-C02-02. G.B.  and C.B. acknowledge financial support from spanish MEC and FEDER (EC) under grant FPA2011-23596,and Generalitat Valenciana under the grant PROMETEO/2008/004.

\appendix

\bibliography{refs}

\end{document}